# Design Choices for Data Governance in Platform Ecosystems – A Contingency Model

**Sung Une Lee**[1,2]
Sungune.Lee@data61.csiro.au

**Liming Zhu**[1,2]
Liming.Zhu@data61.csiro.au

**Ross Jeffery**[1,2]
Ross.Jeffery@data61.csiro.au

[1]Architecture and Analytics Platforms Group, Data61, CSIRO,
Sydney, Australia
[2]School of Computer Science and Engineering, University of New South Wales,
Sydney, Australia

## Abstract

*As platform ecosystems are growing by platform users' data, the importance of data governance has been highlighted. In particular, how to share control and decision rights with platform users are regarded as significant design issues since the role of them is increasing. Platform context should be considered when designing data governance in platform ecosystems (i.e. centralized/decentralized governance). However, there is limited research on this issue. Existing models focus on characteristics for enterprises. This results in limited support for platform ecosystems where there are different types of business context such as open strategies or platform maturity. This paper develops a contingency model for platform ecosystems including distinctive contingency factors. The study then discusses which data governance factors should be carefully considered and strengthened for each contingency in order to succeed in governance and to win market. A case study is performed to validate our model and to show its practical implications.*

# Introduction

Platform ecosystem is recently considered as a key business concept. Sustainable growth of platform ecosystems (e.g. Facebook or Twitter) is enabled through network effects which are based on the interaction of two or more participating groups (Cusumano 2010; Evans 2011; Choudary et al. 2016). As the popularity of platform ecosystems and the value of data in platforms are increasing, the voice of concern about data abuse or misuse is also getting bigger. Therefore, the importance and concern of how to design and implement data governance in platform ecosystems have been highlighted (Lee et al. 2017).

How to partition decision rights and power of a platform ecosystem between a platform owner and platform users becomes a key governance design issue (Tiwana 2013). However, there is limited research on data governance design in this context. Previous research focuses on organizational contingency (e.g. characteristic or strategy of an enterprise) to design an enterprise data governance (Weber et al. 2009). There is a lack of understanding and consideration of platform ecosystems and the underlying complicated relationship caused by multiple participating groups. We have not found a suitable contingency model to support data governance design choices for platform ecosystems. In this study, we discuss two different types of data governance design for platform ecosystems (centralized and decentralized data governance), and provide a contingency model to support decision-making.

The key contributions are threefold. Firstly, this paper provides a deeper understanding of the characteristics and data governance of platform ecosystems. Secondly, we help platform owners' decision-making by introducing design choices (centralized and decentralized governance) and identify contingency factors and the relevant data governance factors that should be considered. Lastly, we demonstrate practical implications of our contingency model and the factors by conducting a case study on an industry platform ecosystem to show how to use the model in the real world.

In the next section, we provide background including platform governance concepts, data governance for platform ecosystems, possible decision choices and related works. We then explain the methodology of this study. Next, we introduce our contingency model and the factors for platform ecosystems, and show the results of a case study. In the last section, we conclude the paper and discuss limitations of the study.

# Background

## *Data Governance Design Choices for Platform Ecosystems*

There are different types of governance such as IT, information, data, platform and social network service (SNS) governance based on the objectives, and which include processes, policies and structures. Traditional IT/data governance support right decision making about IT/data assets within an enterprise (Khatri & Brown 2010). Data governance for platform ecosystems should consider the different business context and concepts even though it shares common characteristics with other types of governance (Dimick 2013; Kamioka et al. 2016). Platform structure, roles, trust, openness and control are considered as key aspects for platform governance design (Schreieck et al. 2016; Hein et al 2016; Tiwana et al. 2010).

A number of research papers have studied governance design choices for platform ecosystems by considering how to share decision rights between platform owners and platform users. Hein et al. (2016) addressed governance structure, which refers to centralized or decentralized governance as critical governance mechanism. The authors claimed that it involves how the authority and responsibility for decision-making is divided among participating groups. Schreieck et al. (2016) described the distribution of power in platform ecosystems. The importance of choices of centralized or decentralized governance is emphasized as platform owners should consider how to balance ownership and power of all sides in the ecosystems (Tiwana 2013; Schreieck et al. 2016) based on platform context (Figure 1 on page 3).

While centralized governance means that a platform owner takes all control power and responsibility, decentralized governance refers to shared governance control and responsibility between a platform owner and platform users. When designing data governance, platform owners should consider which type of design is the better choice based on the characteristics, strategies, situations or future plans of platform ecosystems. We describe the pros and cons of the two types of governance based on a literature review to help platform owners' decision-making (Table 1).

| Perspectives / Design Choices | Platform Owner's Perspective | | Platform User's Perspective | |
|---|---|---|---|---|
| | Pros | Cons | Pros | Cons |
| Centralized | Easy to control desirable behaviors of users and to align business goals and strategies | Slow growing and lots of resources | High quality of data or services (including strong security) | Invisible platforms, lack of trust and hard access to platform ecosystems |
| Decentralized | Save money and effort, increase platform users' satisfaction and participation and fast growing | Hard to control platform ecosystems and users' behaviors and to change goals and strategies | Enhance trust and increase motivation and benefit expectation | Complicated processes, slow decision-making and less secure |

**Table 1. Centralized vs. Decentralized Governance**

## *Design Choices and Data Governance Factors*

It is necessary to think what data governance factors should be considered for designing data governance. Table 2 presents seven data governance factors for platform ecosystems to consider when designing data governance. The factors were identified based on academic literature, and verified and refined through industry governance frameworks and case studies (Facebook, YouTube, EBay and Uber).

| Domain | Factor | Definition | Practice |
|---|---|---|---|
| Data Ownership /Access | Data ownership and access definition | Definition of who owns and uses the data in platform ecosystems. | - Define data ownership of all types of data in the platform (user, process and system data)<br>- Define access rights based on the ownership and contribution of a data contributor |
| | Regulatory environment | Regulations, laws or court cases that could affect the ownership and use of the data in platform ecosystems. | - Identify main criteria for defining data ownership<br>- Consider relevant regulations (laws, standards and cases)<br>- Develop decision models for defining of data owner/access |
| | Contribution measurement | Mechanisms to measure user contribution against value creation by providing data. | - Consider contributors' effort for value creation<br>- Identify dimensions for a measurement model<br>- Combine contribution with data ownership/access definition model |
| Data Usage | Data use case | The purpose of the collected data by platform ecosystems (how to use the data in platform ecosystems). | - Define data categories of a platform including various sources (user, process and system data)<br>- Define data use cases including individual use case based on the data categories<br>- Keep consistency and integrity of the use cases |
| | Conformance | An audit for compliance based on strict processes and rules. | - Recognize requirements for data due processes<br>- Define audit process for conformance of the due processes<br>- Consider the result of audit make visible to stakeholders |
| | Monitoring | Mechanisms to monitor a data supply chain and all activities related to data (including internal activities and external activities by partners or users). | - Detect and notify all activities regarding the use of the data in the platform<br>- Allow all participating groups to monitor and report the use of the data in platforms<br>- Achieve visibility of the data supply chain to stakeholders |
| | Data provenance | Means to trace the derivation history of the data transparently for all participating groups. | - Enable to trace all the derivation history of the data through metadata management<br>- Facilitate data owner authentication through data lifecycle |

**Table 2. Data Governance Factors for Platform Ecosystems (Lee et al. 2017)**

Platform owners need to consider how to place the factors when they design data governance. Some factors (and the practices) may be able to be implemented by platform users (for decentralized governance). In contrast, some factors and the practices perhaps cannot be diffused because of the business characteristics of platform ecosystems and the nature of each factor.

Platform owners can share the control of "contribution measurement", "data provenance", "conformance" and "monitoring" with platform users to enhance trust and visibility of platform ecosystems. Meanwhile, "data ownership and access definition", "regulatory environment" and "data use case" have to be aligned with the business goals and strategies of platform ecosystems. Therefore, the factors cannot be decentralized. Figure 2 shows how the factors can be placed based on business concerns.

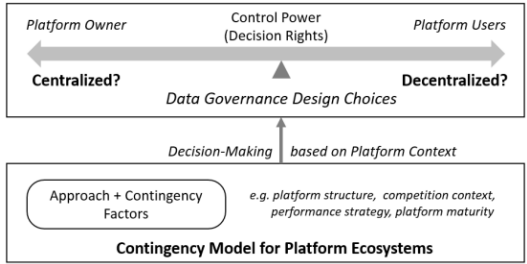 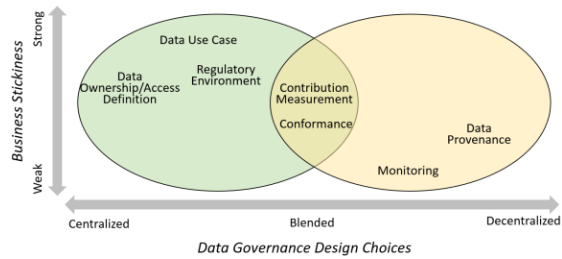

**Figure 1. A Conceptual Model**  **Figure 2. Placement of Data Governance Factors**

### *Prior Research on a Contingency Model for Design Choices*

For better decision-making (toward centralized governance or decentralized governance), it is necessary to look at current and future context or strategy of a platform ecosystem. This issue, however, has received little attention. Weber et al. (2009) emphasized the importance of this, and proposed a contingency model for designing data governance. The authors suggested contingency factors for data governance. The factors are introduced for organizational placement of data quality management activities (DQM). This model is a good source and motivation for different types of governance design. However, the model focuses on the general context for an enterprise. There is a lack of consideration of how it can be applied in a different business context like platform ecosystems. A number of studies provide useful information and ideas for our contingency model for platform ecosystems. A network effect is addressed as a key concept which enables platform's growth and high profit (Weil & Ross 2005; Cusumano 2010; Eisenmann et al. 2006 and 2011, Parker & Van Alstyne 2013 and 2014). Platform governance mechanisms and strategies are mentioned as distinctive factors when there is competition in the market. Parker and Van Alstyne (2013 and 2014) discussed single/multi-homing and open strategy in different competition situations. The authors also paid attention to the level of maturity of a platform ecosystem. They focused on governance styles at the different business stage (depending on the size of market share). Control mechanisms are addressed as key aspects of platform governance (Tiwana 2013; Manner et al. 2013; Schreieck et al. 2016; Hein et al. 2016). Evans (2011) proposed different types of market structures which can be categorized into three platforms based on competition types: coincident, intersecting and monopoly platforms.

In summary, when a platform owner designs data governance, different design choices should be considered based on the context: centralized or decentralized. For this, a contingency model including possible factors should be delivered and carefully reviewed to help a better choice for platform owners. Existing models, however, focus on traditional organizational context, and thus are not explicitly suitable for platform ecosystems. To deal with this, a novel contingency model for platform ecosystems should be provided based on platform ecosystems' characteristics and strategies.

## Methodology

We are inspired by Weber et al.'s contingency model (2009), but we differentiate our suggestion from the model by focusing on platform governance concepts and characteristics. We conducted two stages to develop and validate our model for platform ecosystems (Figure 3). The first stage is to develop a contingency model for platform ecosystems, and the second stage is for validation of the suggested model by performing a case study using a real platform ecosystem.

In the first stage, we developed our model for platform ecosystems including seven factors (Figure 4). While the first and second steps (select and remove) were carried out by analyzing and discussing the referenced model, the third and fourth steps (modify and add) were performed through an academic literature review on platform ecosystems and data governance issues. First of all, we examined Weber et al.'s model to select applicable factors for platform ecosystems (step1. select). We found that "degree of market regulation" is able to be used for platform ecosystems as regulatory environment is also an important data governance factor in using data in platform ecosystems (Lee et al. 2017). In the next step, we removed "process harmonization" because the factor refers to global or local business processes of an enterprise (step2. remove). Next, we looked at the remaining factors, and modified them to adjust to platform ecosystems' situations (step3. modify). Four factors were modified in this step: "performance strategy", "multi-homing strategy", "governance configuration" and "platform market structure". The four factors align the basic concepts of Weber et al.'s model, but they differ in the definition and range of each factor. To differentiate the factors, we reviewed academic literature which addresses platform ecosystem strategy, platform governance concepts and mechanisms, and data governance issues. We identified the distinctive features and situations with respect to data governance in platform ecosystems. We conducted three steps for literature review: keyword searching, backward and forward searching, and literature review based on selection criteria. The main keywords were "((Platform Ecosystem OR multi-sided platform OR two-sided platform) AND (governance OR management))". We also used literature that we searched in our previous studies. The selection criteria was "Include concepts, strategy, or governance mechanisms in the context of platform ecosystem". We excluded the papers that are not related to platform governance, too high level topics (e.g. overview), specific domains and technologies or not academic. In the last step, we found new factors which are not addressed in the referred model (step4. add). This step was also implemented through an extensive literature review. We identified two factors ("open strategy" and "platform maturity") as new factors. The factors are commonly discussed when addressing platform governance in academic literature.

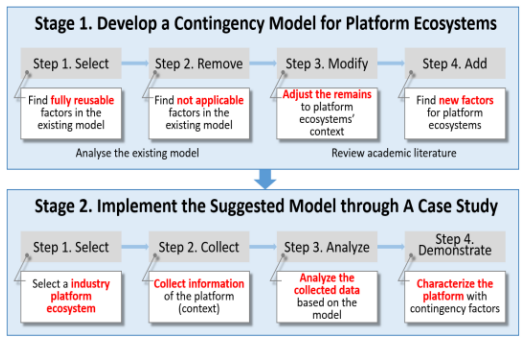
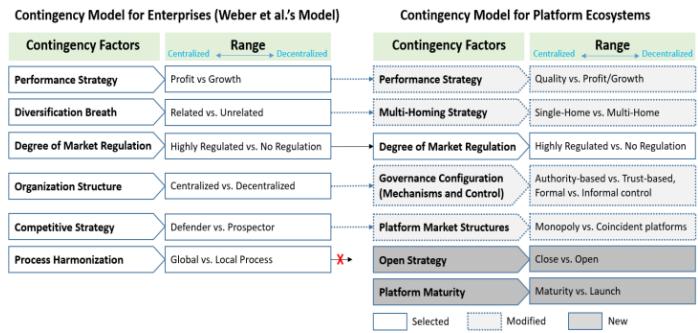

**Figure 3. The Two Stages Approach**          **Figure 4. Contingency Models**

The second stage includes a case study to validate the suggested model and show how to use it in the real world (stage 2 in Figure 3). We chose one industry platform ecosystem, and analyzed the context based on our model (Table 3). We followed Weber et al.'s approach (2009). In the first step, we selected one industry platform ecosystem (Platform A) because one author of this paper worked with the platform in the past (step1. select). Based on the experience of the author, we collected information of Platform A, including current situation (step2. collect). Platform A is a content portal platform. Through the platform, software assets such as software development knowledge, experience or documents are provided and reused by IT companies or individual developers. The platform started the business in 2013 based on government funding to promote the IT industry. Around 3,000 software assets are currently registered in the platform. To encourage user participation, it provides several types of benefits to the participants: e.g. subsidy (technical support, opportunity of advertising or special consulting services) and seeding (best practices, software development guide, UI/UX guide and free software tools). The government also supports the platform by legislating the rule that every SW R&D project funded by the government should register the outcome to the platform. Even with this promotion, Platform A is still struggling to increase growth and to succeed. With the collected information, we analyzed the platform context to show practical implications based on our contingency model (step3. analyze). Lastly, we described the results of our analysis to explain the influence of the contingency factors on the design choices for data governance of Platform A (step4. demonstrate). Thereafter, we identified significant findings to improve our approach and model in future studies.

## Contingency Model for Data Governance in Platform Ecosystems

In this section, we discuss what contingency factors should be considered when platform owners design and implement data governance. We identify seven contingency factors which influence the design choices of data governance in platform ecosystems based on the methodology described in the previous section. Table 3 presents the definition and range of the identified factors and the relevant data governance factors (based on Table 2) in detail.

Design Choices for Data Governance in Platform Ecosystems

## *Contingency Factors*

| Contingency Factor | | Definition (centralized <-> decentralized) | Data Governance Factor | Reference |
|---|---|---|---|---|
| Performance strategy | | Performance objective of a platform ecosystem (profit or growth <-> quality) | Data ownership/access definition, data use case, conformance, monitoring | (Weill & Ross 2005; Weber et al. 2009; Hagiu 2014; Evans 2011) |
| Multi-homing strategy | | Degree of affiliation in a platform ecosystem (single-home <-> multi-home) | Data ownership/access definition, data use case | (Weber et al. 2009; Parker & Van Alstyne 2013 and 2014) |
| Degree of market regulation | | Degree of regulation regarding the use of data in a platform ecosystems (highly regulated <-> no regulation) | regulatory environment, conformance | (Weber et al. 2009; Kaisler et al. 2012 and 2013; Khatri & Brown 2010; Ghazawneh & Henfridsson 2010) |
| Governance Configuration | Governance mechanisms | Type of governance of a platform ecosystem (authority-based <- contract-based -> trust-based governance mechanisms) | Contribution measurement, conformance, monitoring, data provenance | (Tiwana 2013; Manner et al. 2013; De Reuver & Bouwman 2011; Ouchi 1979) |
| | Control mechanisms | Type of control of a platform ecosystem (formal <-> informal control) | Data ownership/access definition, conformance, monitoring, data provenance | |
| Platform market structures | | Type of platform market structures based on competition (coincident <- intersecting -> monopoly platforms) | Data ownership/access definition, data use case, monitoring, data provenance | (Evans 2011; Parker & Van Alstyne 2009 and 2014) |
| Open strategy | | Level of openness of a platform ecosystem (open <-> close) | Data ownership/access definition, data use case, monitoring, data provenance | (Gawer & Henderson 2007; Parker & Van Alstyne 2014; Schreieck et al. 2016; Hein et al. 2016; Choudary et al. 2016) |
| Platform maturity | | Level of maturity of a platform ecosystem (immature <-> mature platforms) | Contribution measurement, monitoring, data provenance | (Cusumano 2010; Parker & Van Alstyne 2014; Schreieck et al. 2016; Hein et al. 2016; Choudary 2013; Choudary et al. 2016) |

**Table 3. Contingency Factors for Data Governance in Platform Ecosystems**

- Performance Strategy: Weill and Ross (2005) addressed profit and growth of organizations to measure IT governance performance. Weber et al. (2009) adopted this concept to data governance contingency. The authors noted that if an organization focuses on profit, the decision-making authority in IT/data governance will be toward centralized. In contrast, if growth is emphasized, the governance will be decentralized. In the context of platform ecosystems, however, profit and growth can be achieved at the same time thanks to network effects of platform ecosystems. As platform ecosystems are growing and increasing revenue based on data uploaded or generated by users, low quality of data issues are also rising (Hagiu 2014). A number of platforms have addressed the issues by introducing formal control processes or monitoring to drive out low quality data (Weber et al. 2009). However, strict control for high quality (through centralized governance) can slow the growth of the platforms.

- Multi-homing Strategy: Some platform owners require their partners to affiliate exclusively with them to offer novel content (single-home). Meanwhile, some platforms allow affiliation with competing platforms to encourage participation: e.g. payment cards in EBay (Parker & Van Alstyne 2014). Platform users can choose one of multiple cards when they purchase products in a platform ecosystem (Evans 2011). Multi-homing is related to openness. If a platform owner chooses "open", less permission (even permissionless) rule to the platform users is followed (Parker & Van Alstyne 2013 and 2014). It means governance should move toward decentralization to share decision rights (Tiwana 2013).

- Degree of Market Regulation: This factor is adopted from the existing model (Weber et al. 2009) as it is consistent with concern of regulatory environment of data governance in platform ecosystems (Lee et al. 2017). The authors noted that a highly regulated environment requires more centralized data governance for compliance. On the contrary, less (or no) regulation might enable platform owners to divide decision rights through a decentralized approach.

- Governance Configuration: Governance mechanisms can be categorized into three categories: authority-based, contract-based and trust-based (De Reuver & Bouwman 2011; Manner et al. 2013). Authority-based mechanisms can be realized within centralized governance by employing the platform owner's power to enforce desirable behavior based on policies. Trust-based mechanisms are used for gaining a certain amount of participants, and then attracting another side platform users. The mechanisms are generally combined with strong incentives to reach the desired goal (Cusumano 2010; Manner et al.

2013). In general, authority-based mechanisms are related to formal control like input, output (outcome) or behavior control, but trust-based mechanisms are enabled by informal control such as clan or social control (Ouchi 1979; Manner et al. 2013) underlying decentralized governance. A contract-based mechanism needs less formality, but still prohibits the unauthorized behavior in using data in platforms.

- Platform Market Structures: Platform market structures (coincident, intersecting, and monopoly platforms) have been addressed under platform competition concerns (Evans 2011). Coincident platforms are recognized when there is too much competition in the market; supply sides (n) > demand sides (m). Monopoly refers to having no competition: n = 1. Coincident platforms (in particular, entrants) might be necessary to be more attractive than incumbent platforms. They tend to be open and share the platforms to encourage users' participation and to win market (Parker & Van Alstyne 2014). Monopoly, however, leans to centralized governance to strictly control the platforms.

- Open Strategy: Open strategy is an important part of the design of a platform ecosystem (Schreieck et al. 2016; Hein et al. 2016). A platform owner opens the platform when there is lack of resources, needs for adoption or innovation (Parker & Van Alstyne 2014). If the degree of openness increases, platform users can easily access the platforms, and platform owners need to share control with users (Schreieck et al. 2016). In consequence, it is toward decentralized. Yet, limited openness leads to high process control, quality and user satisfaction: e.g. Apple's app store (Hein et al. 2016; Choudary et al. 2016).

- Platform Maturity: A newcomer platform has more permissive governance to encourage participation and to reach critical mass when market share is small (Parker & Van Alstyne 2014). To attain enough growth, a platform owner needs less restriction but more trust to attract and lock the users in the platform (Choudary et al. 2016; Schreieck et al. 2016). In particular, trust is regarded as a prerequisite for a platform ecosystem to survive among strong competitors (Schreieck et al. 2016; Hein et al. 2016; Choudary 2013). Thus, less matured platforms tend to prefer multi-homing, open and less negotiation (share control) to capture the benefits of growth. In contrast, when market share is large enough (at maturity), governance has a tendency to be toward tighter control (Parker & Van Alstyne 2014).

### *Contingency Factors and the Relevant Data Governance Factors*

As shown in Table 3, each contingency has a relationship with particular data governance factors. The data governance factors should be focused and enforced based on the linked contingency. For instance, if a platform aims to high quality performance, the platform needs to strengthen data ownership and access control based on clear data use cases (Weill & Woodham 2002; Weill & Ross 2004 and 2005; Khatri & Brown 2010) for strict responsibility and accountability of data quality (Hagiu 2014). Conformance and monitoring mechanisms should be also followed to improve quality performance. Meanwhile, a platform ecosystem which adopts a trust-based governance needs to think about a certain reward for data contributors to encourage participants (Manner et al. 2013; Tang et al. 2012; Chai et al. 2009). In addition, shared conformance, monitoring and data provenance should be considered to achieve visibility of a data supply chain for trust between a platform owner and users (Martin 2015; Ghazawneh & Henfridsson 2010 and 2013). In the same vein, degree of market regulation is strongly related to regulatory environment and conformance to avoid serious court issues (Kaisler et al. 2012 and 2013; Khatri & Brown 2010; Ghazawneh & Henfridsson 2010).

## Case Study

By following the approach in the methodology section (stage 2 in Figure 3), we conducted a case study on the selected industry platform ecosystem (Platform A). We used our model (and the contingency factors in Table 3) to assess and analyze the practical implications. The case study shows that Platform A tends toward centralized governance rather than decentralized. According to the results, the market regulation and platform maturity of the platform lean to decentralized governance, but the other factors show centralized governance trend of the platform.

As shown in Table 4, the suggested contingency model is applicable to analyze and characterize data governance deign of platform ecosystems in practice. The identified factors help a platform owner to systematically analyze their context and to assess the trend of data governance design.

Through this case study, we found two important things to consider for the next steps. First of all, to decide data governance design is not a matter of making a choice between centralized and decentralized (like black and white). Rather, it is like a movable slider between the two design choices (Tiwana 2013). Platform owners should consider the overall context based on the contingency factors, and place it anywhere on the slide bar. To cope with this, each factor in the suggested model should be reviewed and weighted by a platform owner depending on the priority of the factors and business goals. Next, there might be special relationships between the factors as each factor can influence the others. For instance, Platform A has a strong objective of high quality of data (performance strategy). It directly affects two factors (governance configuration and open strategy). As a result, the platform has characteristics of closed and authorized-based control. The two findings should be examined in the next study to generalize and validate our model. The following presents the results of the case study in detail (Table 4).

| Contingency Factor | Context of Platform A | Assessment | Trend (toward) |
|---|---|---|---|
| Performance strategy | - The business goal is reputation and satisfaction of users.<br>- Thus, it aims to high quality of data rather than growth of the platform.<br>* Profit was not considered as the platform is a non-profit platform. | Quality | Centralized |
| Multi-homing strategy | - It tends to prohibit multi-homing for high reputation/differentiation. | Single-home | Centralized |
| Degree of market regulation | - Every data uploaded by users are basically public data based on the prerequisite of the platform policies (except user information).<br>- There is less amount of sensitive data: e.g. personal identifiable information (PII). | No regulation (Less regulation) | **Decentralized** |
| Governance configuration | - There are strict/formal control processes and clear ownership mechanisms by the platform owner. | Authority-based, Formal control | Centralized |
| Platform market structures | - As the platform is supported by the government, the platform dominates the market as a monopoly platform.<br>- There is no competition. | Monopoly | Centralized |
| Open strategy | - It tends to open to demand side users, but still requires login first (only authorized users can access to the platform for the use of data in the platform).<br>- For supply side, the platform is relatively closed as it requires high quality and reputation. | Close | Centralized |
| Platform maturity | - The platform launched four years ago, but it is considered as an immature platform based on the size (the growth rate of registered data and users).<br>- To overcome this, the platform owner adopted subsidy and seeding strategy to attract users. | Immature | **Decentralized** |
| Consequence for data governance design | Based on the assessment above, the data governance design of Platform A tends toward ***"Centralized"***. Platform maturity and degree of market regulation show trend of decentralized governance, but the other factors present it obviously leans to centralized data governance. | | |

**Table 4. The results of a Case Study on Platform A**

# Conclusion and Limitations

In this paper, we discussed design choices for data governance in platform ecosystems, and a contingency model to support decision-making of a platform owner. To demonstrate the practical implications, we carried out a case study by using an industry platform ecosystem.

We conducted a literature review following a lightweight systematic literature review, and developed our contingency model and the seven factors which influence data governance design of platform ecosystems. To differentiate this model from existing models, we focused on the characteristics, strategy, and governance concepts and mechanisms of platform ecosystems. The case study was performed by applying our suggested model and analyzing the context of the selected platform based on each contingency factor. Through the case study, we showed how the model and factors can be used in industry, and designed our future research direction.

This study is currently in progress, and thus remains several limitations. We used only one industry case for the case study. Interviews with the stakeholders of the platform have not yet been carried out. Instead, we simply depended on a qualitative description of the author's working experiences and information. We need to extend this study for validation of our model and observations, and dealing with the findings mentioned in the previous section. In addition, an extensive and a systematic literature review should be considered to justify and refine our suggestion.